\documentclass[twocolumn,amsmath,amssymb]{revtex4}

\newcommand{\commentoutA}[1]{}

\usepackage{graphicx}% Include figure files
\usepackage{dcolumn}% Align table columns on decimal point
\usepackage{bm}% bold math
\begin{document}

\preprint{LA-UR 13-29253}

\title{First principles molecular dynamics without self-consistent field optimization}

\author{Petros  Souvatzis\footnote{Email: petros.souvatsiz@fysik.uu.se}}
\affiliation{Department of Physics and Astronomy, Division of Materials Theory, Uppsala University,
Box 516, SE-75120, Uppsala, Sweden}
\author{Anders M.~N. Niklasson\footnote{Email: amn@lanl.gov}}
\affiliation{Theoretical Division, Los Alamos National Laboratory, Los Alamos, New Mexico 87545, USA}

\date{\today}

%\pacs{71.15.Pd,31.15.Qg, 31.15.Ew}% PACS, the Physics and Astronomy
                             % Classification Scheme.
\begin{abstract}
We present a first principles molecular dynamics approach that is
based on time-reversible extended Lagrangian Born-Oppenheimer molecular dynamics
[Phys. Rev. Lett. {\bf 100}, 123004 (2008)] in the limit of vanishing 
self-consistent field optimization. The optimization-free dynamics keeps
the computational cost to a minimum and typically provides molecular trajectories
that closely follow the exact Born-Oppenheimer potential energy surface. 
Only one single diagonalization and Hamiltonian (or Fockian) construction  
are required in each integration time step.
The proposed dynamics is derived for a general free-energy potential surface
valid at finite electronic temperatures within hybrid density functional theory.
Even in the event of irregular functional behavior that may cause a dynamical instability,
the optimization-free limit represents a natural starting guess for
force calculations that may require a more elaborate iterative electronic
ground state optimization.  Our optimization-free dynamics thus represents a flexible 
theoretical framework for a broad and general class of ab initio molecular dynamics simulations.

\end{abstract}

\maketitle

\section{Introduction}
The molecular dynamics simulation method based on first principles electronic structure theory
is rapidly emerging as a powerful and almost universal tool in materials science, chemistry and molecular biology \cite{DMarx00,BKirchner12}.
A few early applications of Born-Oppenheimer molecular dynamics date back four decades 
ago \cite{MKarplus73,CLeforestier78} and highly efficient density functional methods \cite{hohen,WKohn65}
using plane-wave pseudopotential techniques and the fast Fourier transform \cite{JWCooley65}
appeared already in the mid 80's and early 90's \cite{RCar85,RemlerMadden90,DVanderbilt90,MCPayne92,GKresse93,RNBarnett93} 

One of the major computational obstacles in first principles Born-Oppenheimer molecular dynamics simulations
is the iterative, non-linear, self-consistent field optimization that is required prior to the force calculations 
\cite{RemlerMadden90,PPulay04,DMarx00}.  Many methods have been proposed to overcome this fundamental problem 
\cite{RCar85,MCPayne92,TAArias92,BHartke92,HBSchlegel01,MTuckerman02,HBSchlegel02,JHerbert04,PPulay04,ANiklasson06,TDKuhne07,JAlonso08,JJakowski09}.
One of the more recent approaches is based on an extended Lagrangian formulation of a time-reversible Born-Oppenheimer molecular
dynamics \cite{ANiklasson08,PSteneteg10,GZheng11,JHutter12,LLin13}, which reduces the computational cost of the
self-consistent field optimization while keeping the dynamics stable with respect to long-term energy conservation. 
Extended Lagrangian Born-Oppenheimer molecular dynamics can be used both for metallic and non-metallic
materials \cite{PSteneteg10,LLin13,ANiklasson11},
the integration time step is governed by the slower nuclear degrees of freedom, and it can be used in
combination with fast (non-variational) linear scaling electronic structure solvers without causing a systematic drift in the energy 
\cite{MJCawkwell12}.  It has further been argued that extended Lagrangian Born-Oppenheimer molecular dynamics provides 
a general theoretical framework for alternative forms of first principles molecular dynamics methods \cite{JHutter12}.

The purpose of this paper is to explore some limits of the 
extended Lagrangian formulation of Born-Oppenheimer molecular dynamics. Our main focus is a generalization 
in the limit of vanishing self-consistent field optimization.
This generalization has
previously been investigated within self-consistent-charge tight-binding and Hartree-Fock theory \cite{ANiklasson12,PSouvatzis13}.
In this paper we further extend and explore the optimization-free limit to free-energy potential 
surfaces valid also at finite electronic temperatures within a general hybrid density functional theory.
Our proposed optimization-free dynamics requires
only one diagonalization and in contrast to previous Hartree-Fock simulations \cite{PSouvatzis13} also only one single
effective Hamiltonian construction per time step. This is a significant improvement over the
previous Hartree-Fock calculations, which is made possible by using a particular linearized
expression for the potential free energy. This formulation also provides a computationally
simple force expression that is fully compatible with the potential energy.
For normal simulations that do not encounter irregular (non-convex) behavior in the functional
form around the self-consistent ground state, the optimization-free dynamics yields
trajectories that are practically indistinguishable from an "exact" Born-Oppenheimer
molecular dynamics simulation. However, in the event of anomalous behavior that may
cause numerical instabilities, the optimization-free limit nevertheless also represents a natural
and efficient starting guess for more elaborate force calculations that require an iterative and 
improved accuracy in the electronic ground state optimization or reduced integration time steps 
to recover stability \cite{LLin13,ANiklasson09,PSteneteg10}.  The proposed optimization-free limit of extended Lagrangian 
Born-Oppenheimer molecular dynamics therefore represents a flexible theoretical 
framework for a very broad and general class of materials simulations.

\section{Time-reversible Extended Lagrangian Born-Oppenheimer molecular dynamics}

Extended Lagrangian Born-Oppenheimer molecular dynamics \cite{ANiklasson08,PSteneteg10,GZheng11} enables a time-reversible
integration of the equations of motion that improves the long-term stability of a molecular dynamics simulation, while
keeping the computational cost low by reducing the number of required self-consistent field iterations. 
In our presentation below we will use density matrices for the electronic degrees of freedom, which is a natural
choice for hybrid functionals, i.e. that is easily applicable both in Hartree-Fock \cite{Roothaan, RMcWeeny60} 
and density functional theory \cite{RParr89, RMDreizler90}.
However, the approach is quite general and should be straightforward to apply also to wavefunctions
and the electron density \cite{PSteneteg10,GZheng11}.

In our discussion we will use the term ``ground state'' density matrix and ``Born-Oppenheimer'' molecular dynamics
also for finite temperature ensembles with thermally excited states \cite{RParr89}. We use these terms since 
the self-consistent, fractionally occupied, (i.e. non-idempotent) density matrix minimizes a free energy functional 
that represents a straightforward generalization of regular Born-Oppenheimer molecular dynamics that 
is valid both at zero and finite electronic temperatures \cite{ANiklasson11}.

The finite temperature generalization provides a useful tool for simulations of, for example, metals and
warm dense matter, or simply as an ad hoc tool to avoid self-consistent field instabilities.

\subsection{With self-consistent field optimization}

In time-reversible extended Lagrangian Born-Oppenheimer molecular dynamics the regular dynamical variables for 
the nuclear degrees of freedom are extended with an auxiliary dynamical variable for the electronic degrees of 
freedom, $P$, that evolves close to the optimized self-consistent electronic ground state density matrix, $D$.
The extended Lagrangian equations of motion for the nuclear coordinates, $\{R_I\} = {\bf R}$, for a general
free energy potential surface, $\Omega$, that are valid also at finite electronic temperatures are given by
\begin{equation}\label{XL-BOMD-R}
{\displaystyle M_I{\ddot R}_I = -\frac{\partial \Omega[{\bf R},D]}{\partial R_I}},
\end{equation}
where the dots denote time derivatives.
The equations of motion can be integrated, for example, with the regular velocity Verlet algorithm 
using Hellmann-Feynamn and Pulay forces \cite{RPFeynman39,DMarx00,PPulay69,ANiklasson11}.  
The extended Lagrangian equation of motion for the extended auxiliary 
electronic dynamical variable, $P$, is given by a harmonic oscillator centered around
the self-consistent electronic ground state density matrix, $D$, where 
\begin{equation}\label{XL-BOMD-P}
{\displaystyle {\ddot P} = \omega^2(D - P)}.
\end{equation}
Since $P(t)$ evolves in a harmonic well that follows the ground state, $D(t)$,
the dynamical variable $P(t)$ will stay close to the ground state for sufficiently
large values of the frequency parameter $\omega$ or small integration time steps $\delta t$.
Moreover, since $P(t)$ is a dynamical variable it can be integrated with a time-reversible or
symplectic integration scheme \cite{ANiklasson08,ANiklasson09,AOdell09,AOdell11}. 
In this way we can use $P(t+\delta t)$ as an accurate initial guess to
the self-consistent-field (SCF) optimization procedure of the ground state density matrix, where 
\begin{equation}
D(t+\delta t) = {\rm SCF}[P(t+\delta t)],
\end{equation}
without breaking time reversibility in the underlying electronic degrees of freedom. It is thanks to this
time-reversibility that the long-term conservation of the total energy is stabilized in extended Lagrangian
Born-Oppenheimer molecular dynamics even under approximate and incomplete self-consistent field convergence.

The free energy potential, $\Omega$ in Eq.\ (\ref{XL-BOMD-R}), is here given by
\begin{equation}\label{Free_ES}
\Omega[{\bf R},D] = E[D] -T_e {\cal S}[D] + U^{\rm pair}({\bf R}),
\end{equation}
where $E-T_e {\cal S}$ is
the electronic free energy at electronic temperature $T_e$ with an electronic energy, $E$, and an entropy term, ${\cal S}$. 
$U^{\rm pair}$ is a pair potential term including
ion-ion repulsions and, for example, Van der Waals corrections. The electronic energy term, $E$,
is here assumed to be described by a (restricted) general hybrid density functional expression, with
\begin{equation}\label{simpleE}
E[D] = 2Tr[hD] + Tr[DG^\alpha (D)] + E^{xc}[2D].
\end{equation}
The matrix $h$ corresponds to the one-electron integrals and $G^\alpha(D)$ are the regular Coulomb, $J$,
and exchange, $K$, matrices in Hartree-Fock theory \cite{Roothaan,RMcWeeny60}, 
with the exchange matrix $K$ scaled by a factor $\alpha$ to account for hybrid functionals, i.e.
\begin{equation}
G^\alpha(D) = 2J(D) - \alpha K(D).
\end{equation}
The exchange correlation term $E^{xc}[2D]$ can be a gradient corrected expression \cite{PBE}, a local density approximation \cite{LDA}, or
other mixed functional expressions \cite{B3LYP}, with the electron density given by the (doubly occupied) density matrix, i.e. $2D$.
The ground state density matrix, $D$, which determines the 
potential energy surfaces for the inverse temperature $\beta = 1/(k_B T_e$), 
is given by the self-consistent condition
\begin{equation}\label{Fermi}
D^\perp = \left[e^{\beta\left(H^\perp (D) -\mu_0 I\right)}+1\right]^{-1},
\end{equation}
where the effective single particle Hamiltonian (or Fockian), $H$, is given by
\begin{equation}\label{Heff}
H(D) = h + G^\alpha (D) + V^{xc}(2D).
\end{equation}
The orthogonalized representation of the Hamiltonian, $H^\perp$, in Eq.\ (\ref{Fermi}) above, is calculated through the congruence transformation,
\begin{equation}
H^\perp = Z^T H Z,
\end{equation}
where $Z$ and its transpose $Z^T$ are the inverse factors of the basis set overlap matrix $S$, i.e.
\begin{equation}
Z^TSZ = I,
\end{equation}
and the density matrix in its non-orthogonal form is
\begin{equation}
D = ZD^\perp Z^T .
\end{equation}
The chemical potential, $\mu_0$, is determined to give $D$ the correct occupation of electrons $N_e$, i.e.\ $\mu_0$ is set such that $N_e = 2Tr[D^\perp ]=2Tr[DS]$.
$V^{xc}(2D)$ in Eq.\ (\ref{Heff}) is the regular exchange correlation potential given through the functional derivative of the exchange correlation energy. $V^{xc}(2D)$ is a functional of the electron density given by
the doubly occupied density matrix.
The electronic entropy term ${\cal S}$ in Eq.\ (\ref{Free_ES}) for the spin restricted case, i.e. with double occupation of each orbital, is
\begin{equation}\begin{array}{l}
{\cal S}[D] = -2k_B Tr\left[D^\perp \ln(D^\perp )\right. \\
~\\
~~ + \left. (I-D^\perp ) \ln (I-D^\perp )\right],
\end{array}
\end{equation}
which makes the free energy functional variationally correct at the ground state \cite{MWeinert92,RWentzcovitch92,ANiklasson11}.

\subsection{Without self-consistent field optimization}

The major computational cost of a first principles Born-Oppenheimer molecular dynamics simulation is the 
iterative self-consistent field optimization that is required prior to the force calculations. If a sufficient
degree of optimization is not fulfilled, the Hellmann-Feynman forces are no longer accurate \cite{RPFeynman39,DMarx00,PPulay69,ANiklasson08b},
which in regular Born-Oppenheimer molecular dynamics typically leads to a
systematic drift in the total energy \cite{RemlerMadden90,PPulay04,ANiklasson06}. 
Only by a computationally expensive increase in the
number of self-consistent field iterations is it possible to reduce this drift, though it never fully disappears.
This systematic error accumulation is very unfortunate since the error in each individual
force calculation often is small compared to the local truncation error
caused by the finite integration time step $\delta t$.
The underlying time-reversibility enabled by the extended Lagrangian formulation in Eqs.\ (\ref{XL-BOMD-R})
and (\ref{XL-BOMD-P}) eliminates this problem with respect to the systematic long-term energy drift even
for fairly approximate degrees of self-consistent field optimization. 

In a regular leap-frog or Verlet based integration of the equations of motion, Eqs.\ (\ref{XL-BOMD-R})
and (\ref{XL-BOMD-P}), the local truncations error as measured by the amplitude of the oscillations 
in the total energy (kinetic + potential), scales with the square of the integration time step, i.e. as $\sim \delta t^2$. 
Without a global systematic error accumulation,
the error in time-reversible extended Lagrangian based Born-Oppenheimer molecular dynamics
is thus governed only by the local truncation error $\sim \delta t^2$. 
This gives us the opportunity to further relax 
the accuracy in the individual force calculations as long as any additional error is of the same order as the local
truncation error. This would allow a reduction of the computational cost without
any significant change in the level of accuracy.  As in classical molecular dynamics simulations, the effect 
on the molecular trajectories should be no different than 
using a slightly longer (or shorter) integration time step. What we will demonstrate here, is that this
is possible to achieve, at least under normal simulation conditions, 
even without any self-consistent field optimization at all prior to 
the force calculations. This optimization-free approach keeps the computational cost to a minimum.
Only one single diagonalization and Hamiltonian construction per time step is required.

The equations of motion, Eqs.\ (\ref{XL-BOMD-R}) and (\ref{XL-BOMD-P}),
provide the exact Born-Oppenheimer molecular dynamics only for the exact ground state (gs) density matrix, $D \equiv D_{\rm gs}$.
Since the exact self-consistent ground state in practice never can be reached, even after multiple self-consistent field iterations, 
we always have some small deviation, $\delta D$, from the exact solution, i.e. in real calculations
\begin{equation}
D = Z \left[e^{\beta\left(H^\perp (D_{\rm gs} + \delta D) -\mu_0 I\right)}+1\right]^{-1} Z^T.
\end{equation}
Fortunately, the variational property of the electronic free energy leads to only a minor error and
\begin{equation}\begin{array}{l}
E[D] - T_e{\cal S}[D] = \\
~~\\
~~ E[D_{\rm gs}] - T_e{\cal S}[D_{\rm gs}] + {\cal O}((D-D_{\rm gs})^2).
\end{array}
\end{equation}
Nevertheless, because of the broken commutation between the approximate ground state density matrix $D$ and $H(D)$, 
the Hellmann-Feynman force expression is not valid, which in regular Born-Oppenheimer molecular dynamics 
leads to small but systematic errors in the forces and eventually to a significant loss of long-term accuracy.
Within time-reversible extended Lagrangian Born-Oppenheimer molecular dynamic, 
it is possible to avoid this shortcoming, even without
any self-consistent field optimization at all prior to the force calculations. We achieve this by using a particular 
linearized approximation of the electronic free energy, $E-T_e{\cal S}$.
This linearization provides two important advantages compared to our previous Hartree-Fock simulations \cite{PSouvatzis13}:
(a) only one single Hamiltonian (or Fockian) construction is needed in each time step, and (b) the nuclear forces
are computationally simple yet fully compatible with the linearlized energy expression.

Let the potential free energy be given by the linearized expression
\begin{equation}\label{FreeE}\begin{array}{l}
{\cal F}[P] = 2Tr[hD] + Tr[(2D-P)G^\alpha (P)] \\
~~\\
~~~~ + E^{xc}[2D] - T_e{\cal S}[D],
\end{array}
\end{equation}
where
\begin{equation}\label{DfromP}
D \equiv D(P) = \left[e^{\beta\left(H^\perp (P) -\mu I\right)}+1\right]^{-1}.
\end{equation}
It is then straightforward to show that 
\begin{equation}\label{ErrorOrder}\begin{array}{l}
{\cal F}[P] = E[D_{\rm gs}] - T_e{\cal S}[D_{\rm gs}] \\
~~\\
+ {\cal O}[(D_{\rm gs} -D)^2] + {\cal O}[(D-P)^2].\\
\end{array}
\end{equation}
Moreover, since $P$ is a dynamical variable, we can use a very simple ``Hellmann-Feynman-like'' 
expression for the forces including a basis-set dependent Pulay term that
are given from the partial derivative of the electronic free energy (see Appendix),
\begin{equation}\label{FreeForce}\begin{array}{l}
{\cal F}_R[P] = 2Tr[h_RD] + Tr[(2D-P)G^\alpha_R (P)]\\
~~\\
~~~~ + E^{xc}_R[2D] + 2Tr[S^{-1}H(P)DS_R].
\end{array}
\end{equation}
The last term is the basis-set dependent Pulay term, which here is generalized to finite electronic temperatures
including the electronic entropy contribution \cite{ANiklasson08b}.
The main reason for this simple
form of the force expression, which is valid without any self-consistent optimization of the density matrix $D$, 
is that the partial derivatives are with respect to a constant $P$, 
since it occurs as a dynamical variable. The subscript $R$ in $G^\alpha_R (P)$ and $E^{xc}_R[2D]$ denotes the $R_I$
derivatives with respect to the atomic centered underlying basis set.  
Only one single diagonalization and effective Hamiltonian (or Fockian) construction is needed in each time step. 

The linearized expression of the general hybrid free-energy functional in Eq.\ (\ref{FreeE}) 
and the corresponding forces in Eq.\ (\ref{FreeForce}) represent the underlying theory of this paper.
A detailed derivation starting with Eq.\ (\ref{FreeE}) of the force expression in Eq.\ (\ref{FreeForce}), 
which is valid to second order ${\cal O}[(D-P)^2]$, is given in the appendix.

\subsection{Stability conditions for the electronic integration}

In comparison to exact Born-Oppenheimer molecular dynamics, the 
error in both the forces and the total energy in Eqs.\ (\ref{FreeE}) and (\ref{FreeForce}) 
should be no worse than of order ${\cal O}[(D_{\rm gs}-P)^2] = {\cal O}[(D_{\rm gs} -D)^2] + {\cal O}[(D-P)^2]$,
assuming ${\cal O}[(D_{\rm gs} -D)^2] \le {\cal O}[(D-P)^2]$.
The accuracy of the optimization-free dynamics above is thus governed by an error term ${\cal O}((D_{\rm gs}-P)^2)$.
It is therefore important to keep the auxiliary dynamical variable $P$ as close as possible to the ground state.
Since $P$ only moves towards the exact ground state, $D_{\rm gs}$, through the harmonic oscillator centered around the approximate ground state $D$,
the equation of motion for the electronic degrees of freedom, Eq.\ (\ref{XL-BOMD-P}), which formally is derived
within the extended Lagrangian framework for $D \equiv D_{\rm gs}$, will in general be unstable unless
\begin{equation}
\|D-D_{\rm gs}\| < \|P-D_{\rm gs}\|. 
\end{equation}
To enable stability in our pursued limit of only one single diagonalization per time step, 
we may use an approximate equations of motion for the electronic degrees of freedom,
\begin{equation}
{\ddot P} = \omega^2({\widetilde D} - P),
\end{equation}
where ${\widetilde D}$ is some improved approximation, compared to $D$, of the exact ground state, i.e.
\begin{equation}
{\widetilde D} = {\widetilde D}(D,P) \approx D_{\rm gs},
\end{equation}
Possibly the simplest choice is a linear mixing where
\begin{equation}\label{LinMix}
{\widetilde D}(D,P) = \gamma D + (1-\gamma )P.
\end{equation}
This choice leads to the same form for the equations of motion as in (SCF-optimized) 
extended Lagrangian Born-Oppenheimer molecular dynamics,
\begin{equation}
{\ddot P} = \gamma \omega^2(D - P),
\end{equation}
but now with $\omega^2$ scaled by a constant $\gamma$. In this case, stability can 
be achieved whenever the functional form of ${\cal F}[P]$ is convex in the sense 
that there exist a constant $\gamma  \in [0,1]$ such that 
\begin{equation}
{\cal F}[\gamma D + (1-\gamma )P] < \gamma {\cal F}(D) + (1-\gamma) {\cal F}(P). 
\end{equation}
In this simple case of linear mixing, which we have used throughout all our
calculations, we can therefore use the regular unoptimized equation of motion in Eq.\ (\ref{XL-BOMD-P}), with $\Omega$
defined through ${\cal F}[P]$ in Eq.\ (\ref{FreeE}) and with $\omega^2$ 
rescaled by a factor $\gamma  \in [0,1]$.  This approach is simple and straightforward and works for normal convex functional forms. 
In the (rare) event of functional anomalies, for example, due to broken functional convexity with a self-consistent field instability, we may 
have to adjust the scaling factor $\gamma$, reduce the integration time step $\delta t$, use more advanced (preconditioned) 
approximations for ${\widetilde D}(D,P)$, introducing an ad hoc electronic thermal smearing
or revert to a more costly iterative self-consistent field optimization procedure.
The automatic ``on-the-fly'' detection and development of such adaptive integration methods is an 
area for future research.

An interesting analysis of stability and accuracy conditions of the equations of motion derived from 
extended Lagrangian Born-Oppenheimer molecular dynamics can be found in a recent paper by Lin et al. \cite{LLin13}.
This paper also includes a detailed comparison between time-reversible extended Lagrangian
Born-Oppenheimer molecular dynamics and the more well-known method by Car and Parrinello \cite{RCar85}. 
A key difference is their sensitivity to the electronic gap and thus the ability to simulate 
metallic systems, which is straightforward with extended Lagrangian Born-Oppenheimer molecular dynamics \cite{PSteneteg10}. 
Another difference is that the constant of motion in Car-Parrinello 
molecular dynamics includes the electronic kinetic energy. This is not the case in extended Lagrangian
Born-Oppenheimer molecular dynamics, which is formally derived in the adiabatic limit of vanishing (zero) 
electron mass. A generalized presentation of both Car-Parrinello and extended
Lagrangian Born-Oppenheimer molecular dynamics is also given in a recent paper by Hutter \cite{JHutter12}.

\subsection{Equations of motion for a fast quantum mechanical molecular dynamics}

As a summary of the discussion above we have derived, based on extended Lagrangian Born-Oppenheimer 
molecular dynamics, a fast quantum mechanical molecular dynamics (fast-QMMD) scheme
that requires only one single diagonalization and Hamiltonian construction per time step. The nuclear
degrees of freedom is governed by
\begin{equation}\label{Fast-QMMD-R}
{\displaystyle M_I{\ddot R}_I = -\left.\frac{\partial \Omega[{\bf R},P]}{\partial R_I}\right|_P},
\end{equation}
and the electronic evolution by
\begin{equation}\label{Fast-QMMD-P}
{\ddot P} = \omega^2\left({\widetilde D} - P\right) = \gamma \omega^2\left(D - P\right).
\end{equation}
The nuclear forces in Eq.\ (\ref{Fast-QMMD-R}) are given for the partial derivatives of $\Omega$ with
respect to nuclear coordinates $R_I$ under the condition of constant $P$.
The potential energy is
\begin{equation}\label{Fast-QMMD-FreeE}\begin{array}{l}
{\Omega}[{\bf R},P] = 2Tr[hD] + Tr[(2D-P)G^\alpha (P)]\\
~~\\
~~ + E^{xc}[2D] - T_e{\cal S}[D] + U^{\rm pair}({\bf R}),
\end{array}
\end{equation}
with the corresponding forces
\begin{equation}\label{Fast-QMMD-Force}\begin{array}{l}
{\Omega}_R[{\bf R},P] = 2Tr[h_RD] + Tr[(2D-P)G^\alpha_R (P)]\\
~~\\
~~+ E^{xc}_R[2D] + 2Tr[S^{-1}H(P)DS_R] + U^{\rm pair}_R({\bf R}).
\end{array}
\end{equation}
The density matrix $D$ at $T_e \ge 0$ is
\begin{equation}\label{Fast-QMMD-D}
D \equiv D(P) = Z\left[e^{\beta\left(H^\perp (P)-\mu I\right)}+1\right]^{-1}Z^T,
\end{equation}
where
\begin{equation}\label{Fast-QMMD-H}
H^\perp (P) = Z^T\left(h + G^\alpha (P) + V^{xc}(2P)\right)Z.
\end{equation}
As above, the congruence factor $Z$ and its transpose $Z^T$ are determined by
$Z^TSZ = I$, where $S$ is the basis-set overlap matrix.
We integrate the nuclear degrees of freedom with a regular velocity-Verlet scheme. The electronic
equation of motion, Eq. (\ref{Fast-QMMD-P}), is integrated with a modified Verlet scheme \cite{ANiklasson09,PSteneteg10,GZheng11}
that removes a possible accumulation numerical noise through a weak dissipation.
For the linear mixing approximation of ${\widetilde D}$ in Eq.\ (\ref{LinMix}), this modified Verlet integration is 
\begin{equation}\label{VRL_Damp}
P_{n+1} = 2P_{n} - P_{n-1} + \gamma \kappa (D_{n} - P_{n}) + \alpha \sum_{k = 0}^{K} c_k P_{n-k},
\end{equation}
where the dimensionless variable $\kappa = \delta t^2 \omega^2$ is rescaled by the constant $\gamma \in [0,1]$.
Three material independent sets of optimized coefficients for $\alpha$ and $c_k$ are given in Tab. \ref{Tab_Coef}.
Alternative higher-order symplectic integration schemes can also be applied \cite{ANiklasson08,AOdell09,AOdell11}, 
but have not been used in the present study.

\begin{table}[t]
  \centering
  \caption{\protect Coefficients for the Verlet integration scheme with the external dissipative
  force term in Eq.\ (\ref{VRL_Damp}). The coefficients are derived in Ref.\ \cite{ANiklasson09}, which
  contains a more complete set of coefficients.
  }\label{Tab_Coef}
  \begin{ruledtabular}
  \begin{tabular}{llccccccccc}
    $K$ & $\delta t^2 \omega^2$ & \!\!$\alpha\!\times\!10^{-3}$\!\! & \!$c_{0}$ & $c_{1}$ & $c_{2}$ & $c_{3}$ & $c_{4}$ &
$c_{5}$ & $c_{6}$ & $c_{7}$ \\
     \hline
     5  & 1.82 & \!\!18\!\! &-6    &   14   &  -8  & -3  &  4  &  -1  &     &      \\
     6  & 1.84 & \!\!5.5\!\! &-14    &   36   &  -27  & -2  &  12  &  -6  &  1  &      \\
     7  & 1.86 & \!\!1.6\!\! &-36    &   99  & -88  & 11  &  32 & -25  & 8   &  -1  \\
  \end{tabular}
  \end{ruledtabular}
\end{table}

\section{Examples and Analysis}

The proposed fast-QMMD scheme, Eqs.\ (\ref{Fast-QMMD-R} - \ref{VRL_Damp}), has been implemented in
the Uppsala Quantum Chemistry code (UQuantChem) \cite{UQuantChem}, which is a freely available program 
package for ab initio electronic structure calculations using Gaussian basis set representations that
can be obtained from the code's homepage \cite{code}.  
The UQuantChem program includes Hartree-Fock theory, diffusion and variational Monte Carlo, configuration interaction, 
Moller-Plesset perturbation theory, density functional theory and hybrid functionals, finite
temperature calculations, structural optimization and first principles molecular dynamics simulations.
Our fast-QMMD implementation in the UQuantChem package has been
carefully commented in reference to this paper in order to facilitate the implementation of the 
fast-QMMD scheme into a broader variety of electronic structure software. Notice, that details of
the implementation are very important and it may not always be straightforward to implement
the fast-QMMD scheme in a regular first principles electronic structure code.

All our ``exact'' Born-Oppenheimer molecular dynamics (BOMD) simulations were performed based on 
extended Lagrangian Born-Oppenheimer molecular dynamics, Eqs.\ (\ref{XL-BOMD-R}) and (\ref{XL-BOMD-P}),
using 5 self-consistent-field optimization cycles per time step prior to the force calculations.
This choice provides a convergence in the potential free energy of about $\sim 10^{-7}$ Hartree, which
is 2-3 orders of magnitude smaller than the oscillations in the total energy that are used in the
comparisons.

\subsection{Comparison to Born-Oppenheimer molecular dynamics}

\begin{figure}[tbp]
\includegraphics*[angle=0,scale=0.3]{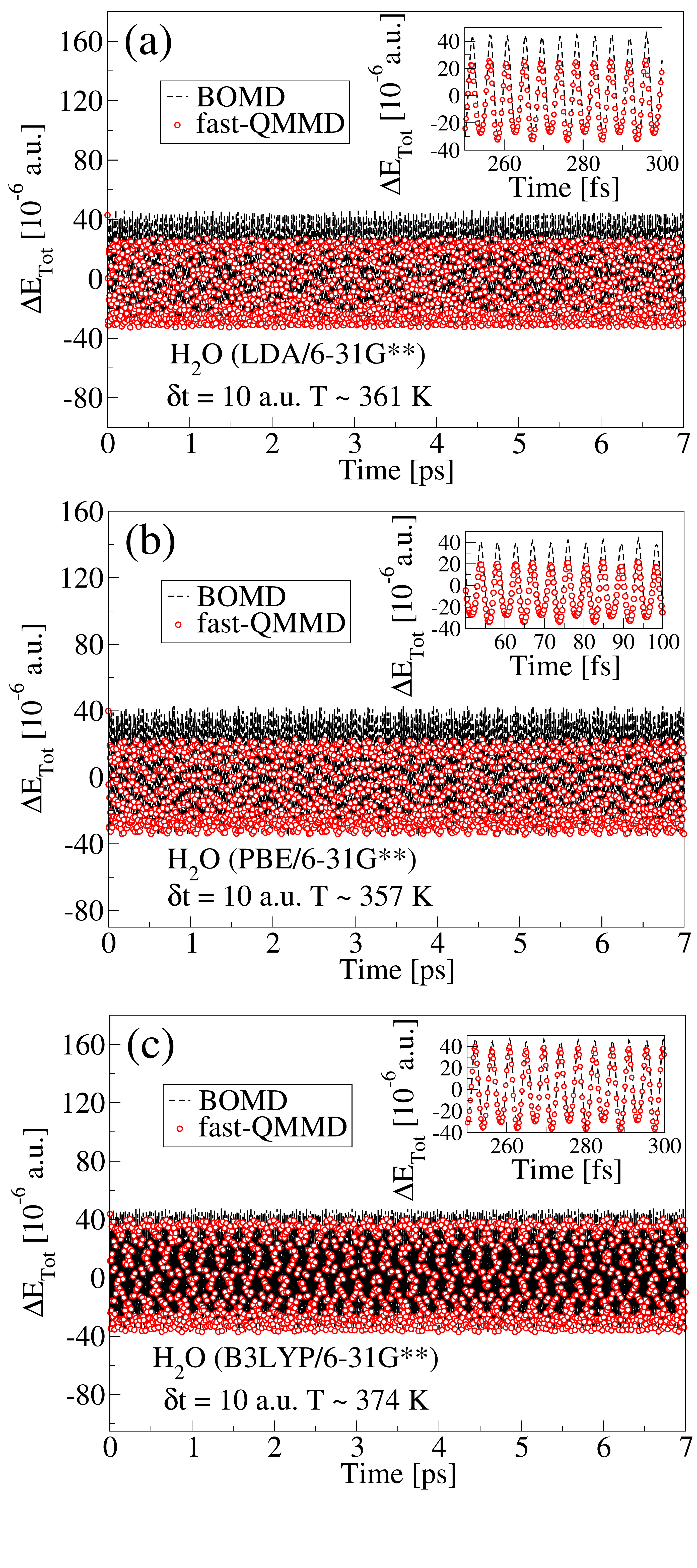}
\caption{\label{Figure_1}\small
Total energy fluctuations (kinetic + potential) for water using
self-consistent Born-Oppenheimer molecular dynamics (BOMD),
and the first principles fast-QMMD, Eq.\ (\ref{Fast-QMMD-R}-\ref{VRL_Damp}).
For the fast-QMMD simulation we used the electronic free energy expression in Eq.\ (\ref{Fast-QMMD-FreeE})
in the total energy estimates.
%In (a)  the LDA density functional \cite{LDA}  was used, here  $E_{0}$ = -75.84438 a.u.
%In (b)  the PBE \cite{PBE} density functional  was used, here $E_{0}$ = -76.38535 a.u.
%In ({c}) the hybrid B3LYP \cite{B3LYP} density functional  was used, here $E_{0}$ = -76.37234 a.u.
All calculations were performed with the modified Verlet integration ($K$ = 7 in table \ref{Tab_Coef}) and 
rescaling of $\kappa = \delta t^2 \omega^2$ with a factor $\gamma$ = 0.7.
The time step used was 10 a.u.}
\end{figure}

In Fig.\ \ref{Figure_1} we show a comparison between our fast-QMMD, 
Eqs.\ (\ref{Fast-QMMD-R}-\ref{VRL_Damp}), and self-consistent Born-Oppenheimer molecular dynamics with respect
to the fluctuations in the total energy (kinetic + potential), i.e. the local truncation error, for simulations
of a single water molecule. We have used three different levels of density functional theory:  a)
the local density approximation (LDA) \cite{LDA}, b) the gradient corrected approximation (PBE) \cite{PBE}, and
c) a hybrid functional (B3LYP) \cite{B3LYP}. In all cases we find that the total energy fluctuations,
\begin{equation}
E_{\rm Tot}(t) = \frac{1}{2}\sum_I M_I{\dot R}^2_I(t) + \Omega({\bf R},t),
\end{equation}
behave in a very similar way.
For the fast-QMMD simulations we used the potential energy expression of $\Omega({\bf R},t)$ in Eq.\ (\ref{Fast-QMMD-FreeE})
to estimate the total energy.

In the next figure, Fig.\ \ref{Figure_2} (a), we show interatomic distances between self-consistent Born-Oppenheimer
molecular dynamics and our proposed fast-QMMD scheme for a C$_2$H$_6$ molecule using the local density approximation. 
The upper panel shows virtually no difference
between the fast optimization-free scheme and the optimized ``exact'' Born-Oppenheimer simulation.
The curves are essentially on top of each other even after 500 fs of simulation time. For any chaotic 
dynamical system like this, the curves will eventually diverge due to any infinitesimal
perturbation. The lower panel, Fig.\ \ref{Figure_2} (b), shows an example of the interatomic C-F distance in a CF$_4$
molecule. Here we use a four times longer integration time step, $\delta t$ = 40 a.u., and we 
find a small shift in the frequency that is visible after a few hundred time step.
For the first 100-200 fs of simulation time the curves are very close.

\begin{figure}[tbp]
\includegraphics*[angle=0,scale=0.3]{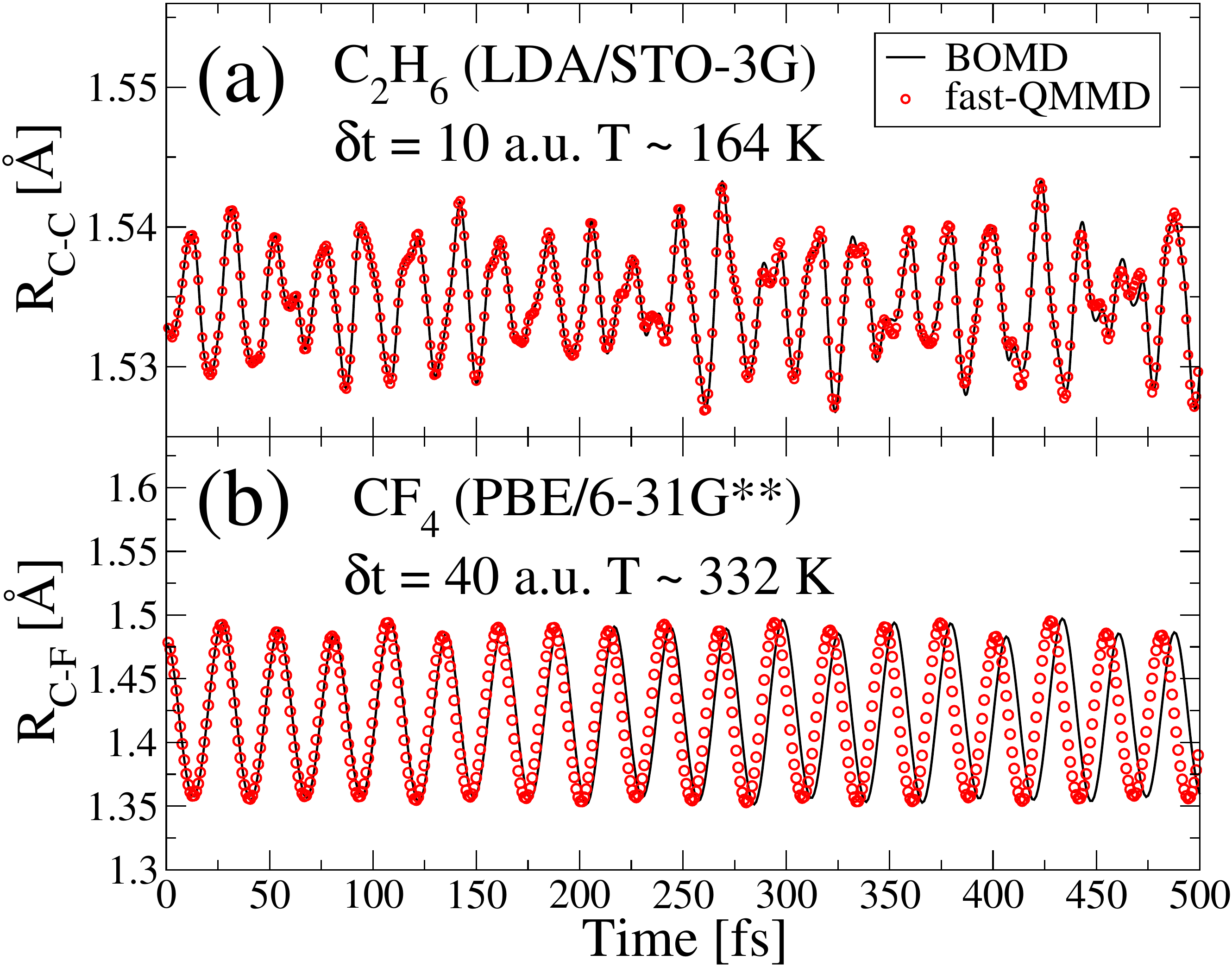}
\caption{\label{Figure_2}\small
Interatomic distances calculated using
self-consistent Born-Oppenheimer molecular dynamics (BOMD), 
and the first principles fast-QMMD, Eq.\ (\ref{Fast-QMMD-R}-\ref{VRL_Damp}).  In (a), the interatomic distance between the Carbon atoms
, R$_{C-C}$, in a C$_{2}$H$_{6}$ molecule  is displayed. In (b), the interatomic distance between the Carbon atom 
and one of the Fluorine atoms, R$_{C-F}$, in a CF$_{4}$ molecule  is displayed.
In (a)  the LDA density functional \cite{LDA} together with  a STO-3G basis-set  and  a 10 a.u. time step was used. In (b)  the PBE density functional \cite{PBE} together with  a  6-31G$^{**}$ basis-set and  a 40 a.u. time step was used.
The ethan calculations were performed with a modified Verlet integration ($K$ = 7 in table \ref{Tab_Coef} ) and rescaling of $\kappa = \delta t^2 \omega^2$ with a factor $\gamma$ = 0.7.
The CF$_{4}$ calculations were performed with a modified Verlet integration ($K$ = 7 in table \ref{Tab_Coef} ) and rescaling of $\kappa = \delta t^2 \omega^2$ with a factor $\gamma$ = 0.54. }
\end{figure}

\begin{figure}[tbp]
\includegraphics*[angle=0,scale=0.30]{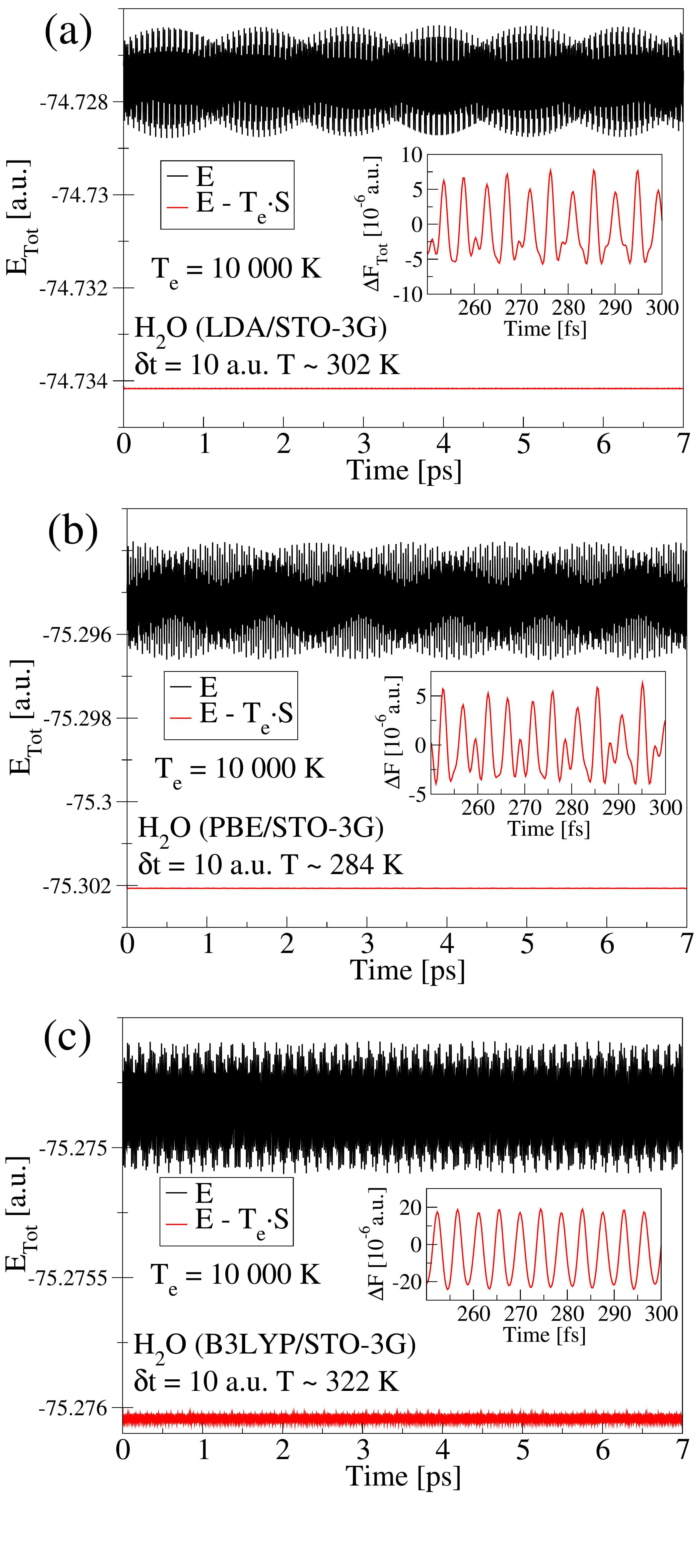}
\caption{\label{Figure_3}\small
Total energy including internal energy ($E$) and 
free energy ($E-TS$) for water using self-consistent 
Born-Oppenheimer molecular dynamics (BOMD),
and the first principles fast-QMMD, Eq.\ (\ref{Fast-QMMD-R}-\ref{Fast-QMMD-P}).
In (a)  the LDA density functional \cite{LDA}  was used, 
in (b)  the PBE \cite{PBE} density functional  was used and 
in ({c}) the hybrid B3LYP \cite{B3LYP} density functional  was used. All the  calculation was  performed with the 
modified Verlet integration (using $K$ = 7 in table \ref{Tab_Coef} ) and a rescaling of $\kappa = \delta t^2 \omega^2$ using a factor $\gamma$ = 0.7.}
\end{figure}

\begin{figure}[tbp]
\includegraphics*[angle=0,scale=0.30]{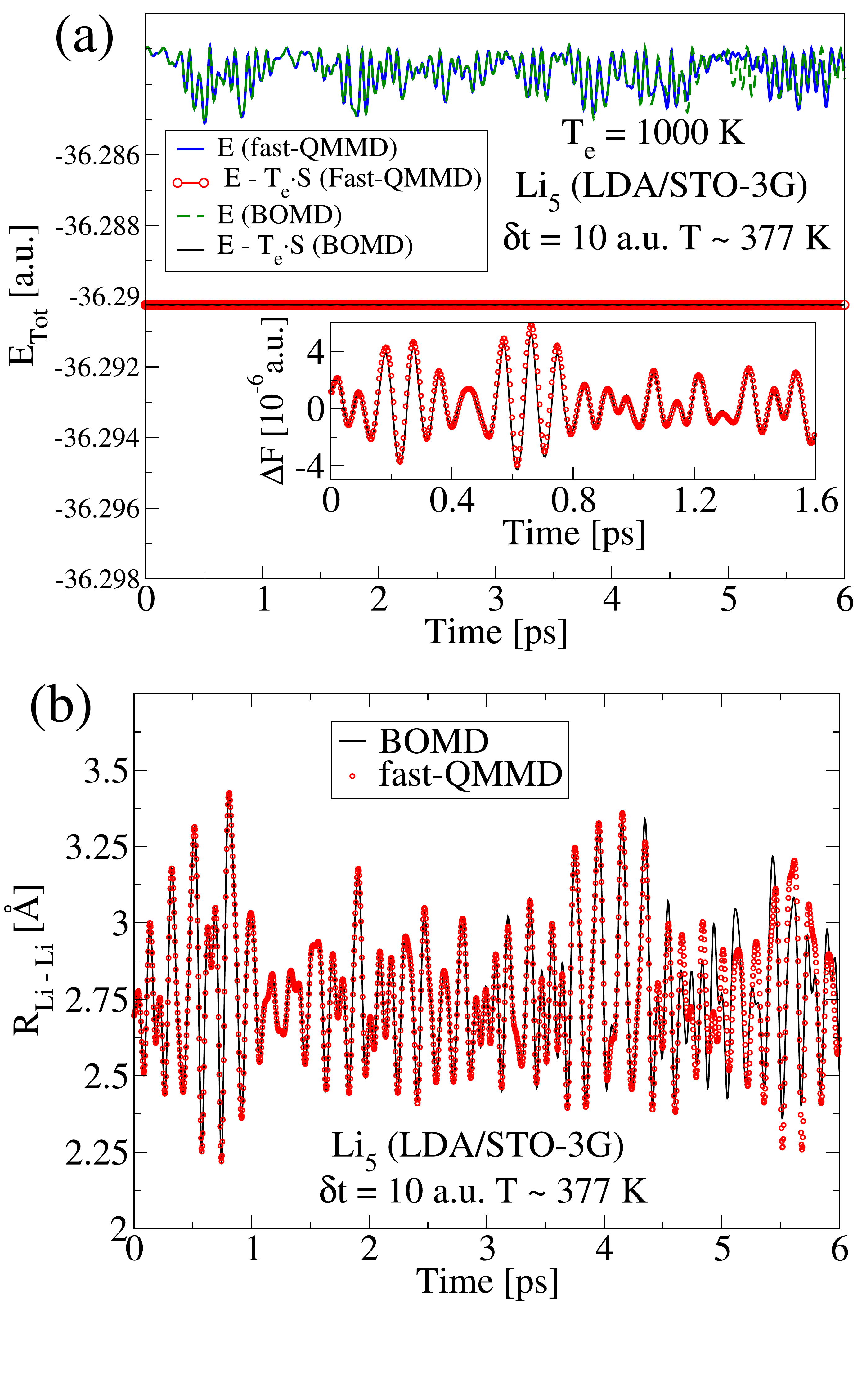}
\caption{\label{Figure_4}\small
In (a) the total internal and free energy of a fast-QMMD calculation of a small Lithium cluster (Li$_{5}$) 
with a 0.62 eV energy band-gap using the LDA density functional \cite{LDA}, here 
compared to the corresponding self-consistent Born-Oppenheimer molecular dynamics (BOMD) calculation. In (b), the distance between two Lithium atoms 
calculated with the fast-QMMD scheme is compared to the same distances obtained with the Born-Oppenheimer molecular dynamics (BOMD) calculation.
The the integrations were performed with a time step of 10 a.u. using a modified Verlet scheme (with $K$ = 5 in table \ref{Tab_Coef} ) and rescaling of a 
$\kappa = \delta t^2 \omega^2$ with a factor $\gamma$ = 0.27.}
\end{figure}

\begin{figure}[tbp]
\includegraphics*[angle=0,scale=0.30]{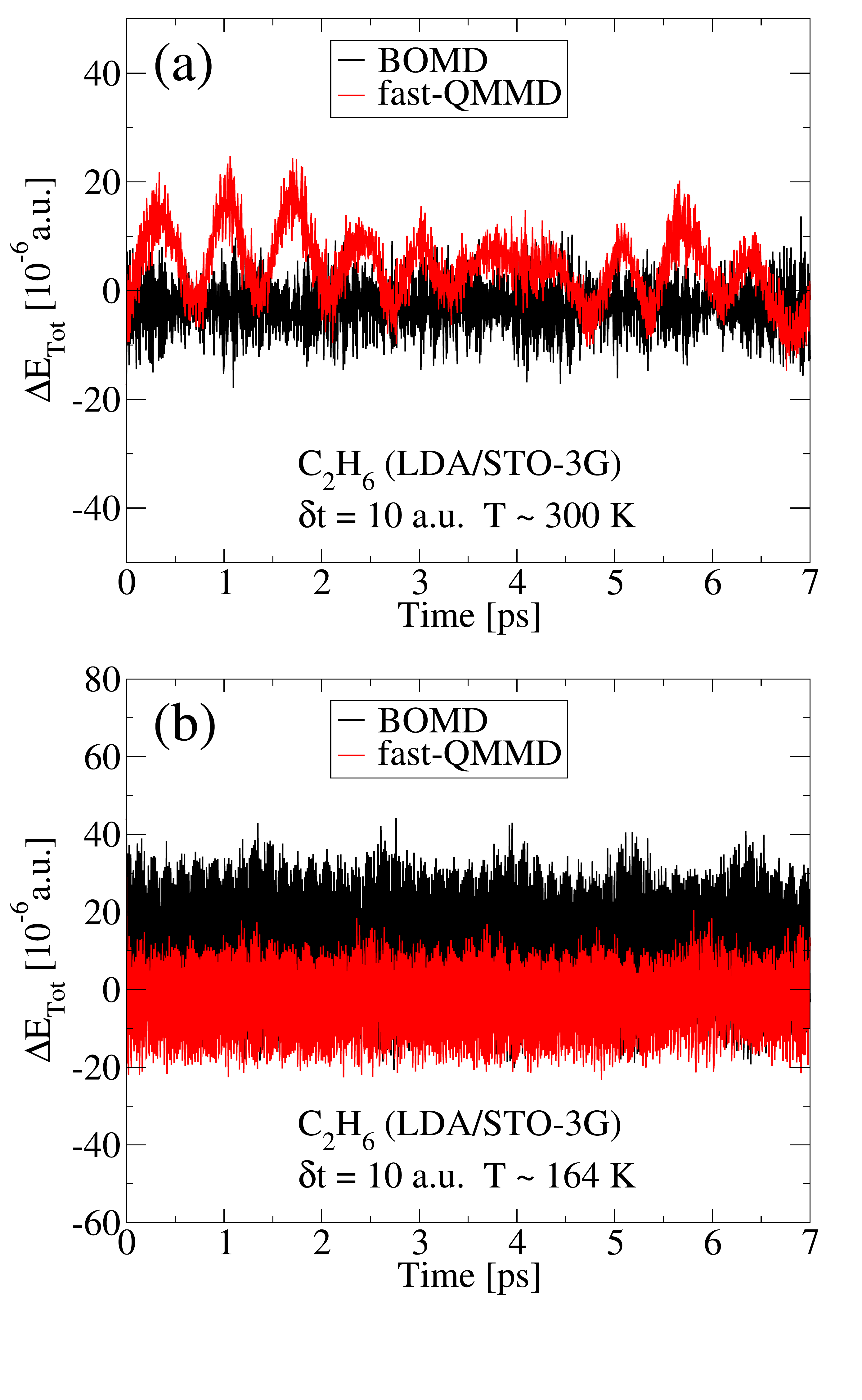}
\caption{\label{Figure_5}\small
Total energy fluctuations for C$_{2}$H$_{6}$ for rotational and vibrational dynamics in (a) and pure vibrational dynamics in (b), using
self-consistent Born-Oppenheimer molecular dynamics (BOMD),
and the first principles fast-QMMD, Eq.\ (\ref{Fast-QMMD-R}-\ref{Fast-QMMD-P}) utilizing the LDA density functional \cite{LDA}.
 %Here  $E_{0}$ = -78.06437 a.u. in (a) and $E_{0}$ = -78.07521 a.u in (b). 
 All calculations were 
The integrations were performed with a time step of 10 a.u. using the modified Verlet algorithm (with $K$ = 7 in table \ref{Tab_Coef} ),  
and a rescaling of $\kappa = \delta t^2 \omega^2$ with a factor $\gamma$ = 0.7.}
\end{figure}

\begin{figure}[tbp]
\includegraphics*[angle=0,scale=0.3]{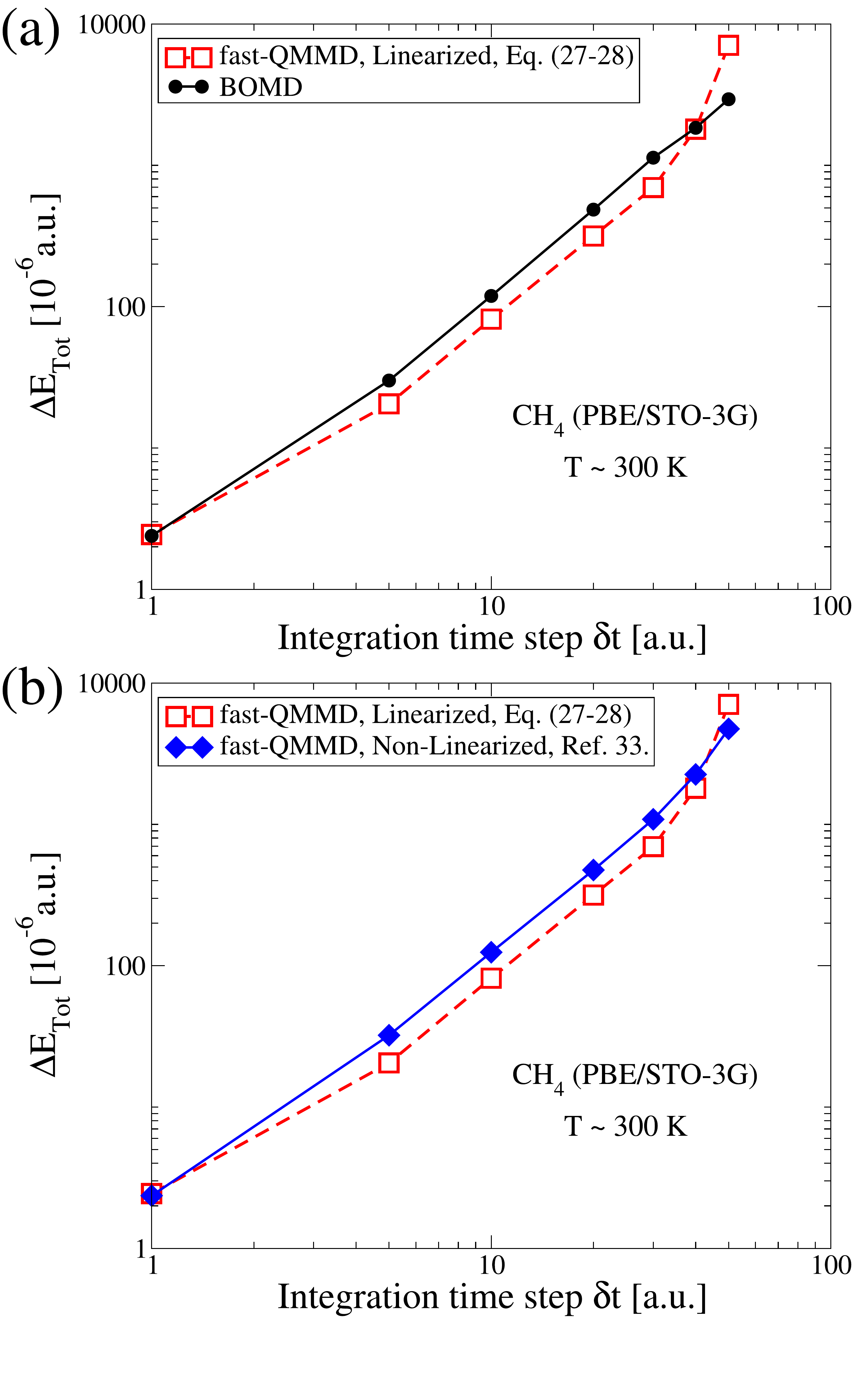}
\caption{\label{Figure_6}\small
The local truncation error in a fast-QMMD calculation utilizing the linearized energy expression, 
Eq.\ (\ref{Fast-QMMD-FreeE}), and force expression, Eq.\ (\ref{Fast-QMMD-Force}),  for vibrational motion
of Methane for different time-steps.
In (a) compared to the corresponding ``exact'' (5 SCF/step) Born-Oppenheimer molecular dynamics (BOMD)
calculations, and in (b) compared  to a fast-QMMD calculation using the corresponding ``non-linearized''
optimization-free energy and force expressions used in the previous study \cite{PSouvatzis13}.
 Here the PBE functional \cite{PBE} was used together with a STO-3G basis-set. 
All calculations were performed with a modified Verlet scheme
(using $K$ = 7 in table \ref{Tab_Coef} ) and a rescaling of $\kappa = \delta t^2 \omega^2$ with a factor $\gamma$ = 0.7. }
\end{figure}

\subsection{Finite electronic temperatures}

Figure \ref{Figure_3} illustrates the total energy conservation with ($E-T_eS$) and without ($E$) the electronic entropy contribution.
At an electronic temperature of 10,000 K the fluctuations in total energy decrease by 1-2 orders of magnitude when the
electronic entropy contribution is included. This high
resolution could be critical in order to resolve small differences in the free energy between competing phases.
Here it mainly illustrates that the constant of motion behaves as expected. 

Because of the large HOMO-LUMO gap of the water system in Fig.\ \ref{Figure_3}, a fairly high temperature
is needed to illustrate the electronic temperature effect. In Fig.\ \ref{Figure_4} we show the corresponding
finite temperature result for a small Lithium cluster (Li$_5$). In the restricted calculation of
a system with an odd number of electrons, the highest occupied state has an occupation factor of $\sim 1/2$ and
the entropy contribution is thus significant even at a fairly low electronic temperature ($T_e = 1000$ K), as
is seen in the upper panel (a).  This example also illustrated the ability of the fast-QMMD scheme to simulate 
systems that may have significant problems to reach the self-consistent ground state. The lower panel
shows the interatomic distance between two Li atoms. The fast-QMMD scheme provides molecular trajectories
that are virtually on top of the Born-Oppenheimer result for over 4 ps of simulation time.

\subsection{Translation, vibration, and rotation}

It may appear that the fast-QMMD scheme is almost identical to an ``exact'' fully converged Born-Oppenheimer
simulation. However, there is a subtle difference in the behavior of the local truncation error.
Consider a pure translational motion of a molecular system. In this case there will be no transfer of
energy from the kinetic to the potential energy. Given an atom centered basis set, the electronic ground state
density matrix will remain constant and the local truncation error, i.e. the error in the ability to account
for the correct energy transfer between kinetic and potential energy, will effectively be zero, both
for a fast-QMMD and a Born-Oppenheimer simulation.  This is not the case for a vibrational motion, where
both fast-QMMD and Born-Oppenheimer simulations will have local truncation errors due to the finite
integration time step and fail to exactly account for the balance in energy transfer between kinetic
and potential energy. However, for rotational motion a difference appears between fast-QMMD and Born-Oppenheimer 
molecular dynamics simulations.  As for a pure translational motion, a pure rotational mode has no energy 
transfer of kinetic energy. Nevertheless, since the matrix representation of the ground state density matrix change 
between time steps for a rotational motion, the fast-QMMD will show some small changes in the total energy
due to the rotation, which will be absent in an ``exact'' Born-Oppenheimer simulation. Figure \ref{Figure_5}
illustrates this behavior.  In the upper panel (a), which has a superposed vibrational and rotational motion, 
we find an additional oscillatory motion in the total energy
of the fast-QMMD simulation compared to the Born-Oppenheimer simulation. This qualitative difference is not found
in the lower panel (b), which only has a pure vibrational motion. For the general case of composite motion, 
we therefore expect fast-QMMD simulations to have a slightly increased local truncation error compared to
an ``exact'' fully converged Born-Oppenheimer simulation.

\subsection{Tunable accuracy}

The sensitivity to the integration time step $\delta t$ could potentially be a limiting factor for
the fast-QMMD scheme.  Figure \ref{Figure_6} (a) shows the local truncation error (measured by 
the amplitude of the oscillations of the total energy) as a function of the length of 
the time step $\delta t$ for fast-QMMD and Born-Oppenheimer simulations of a single
Methane molecule at room temperature. The two curves both scale as $\sim \delta t^2$, where the  
fast-QMMD truncation errors are slightly smaller than the BOMD errors all the way up to $\delta t \approx 40$ a.u. (about 1 fs), 
where a small shift appears.  Just as in a classical molecular dynamics simulation, 
the local truncation error is governed by the size of a {\em tunable} integration time step and 
not by the number of self-consistent field iterations. The general validity of this result is hard to judge, 
but the behavior in Fig.\ \ref{Figure_6} (a) is consistent with what we have found so far 
in our simulations and can be expected to hold as long as there are no inherent self-consistent 
field instabilities or anomalous (non-convex) functional behavior.

\subsection{Comparison to previous approach}

In Fig.\ \ref{Figure_6} (b) we show a comparison of the local truncation error as a function of the integration time step, $\delta t$, 
between our fast-QMMD using the linearized energy and force expression, Eqs.\ (\ref{Fast-QMMD-R}-\ref{VRL_Damp}), 
and the corresponding fast-QMMD based on a ``non-linearized'' potential energy that was used in our previous Hartree-Fock simulations \cite{PSouvatzis13}.
Both methods require only one diagonalization per time step, but in contrast to the new linearized scheme, the previous method requires
two full constructions of the effective Hamiltonian in each time step and the forces are approximate. The linearized version of the fast-QMMD scheme yields slightly smaller 
local truncation errors compared to the non-linearized version all the way up to a time-step, $\delta t \sim$ 40 a.u. ($\sim 1$ fs) above which the linearized
formulation starts to diverge. We believe this improved behavior can be explained by the consistency between the forces and the potential
energy for the linearized expression, which is valid as long as the linearization is sufficiently accurate, i.e. for $\delta t \lesssim$ 40 a.u.
For normal simulations, with time steps of about 10-30 a.u., the stability and accuracy of the two approaches are comparable, but thanks
to the lower complexity of the linearized approach, with only one Hamiltonian construction per time step, our new scheme represents a significant
improvement.

\subsection{Stability}

\begin{figure}[tbp]
\includegraphics*[angle=0,scale=0.3]{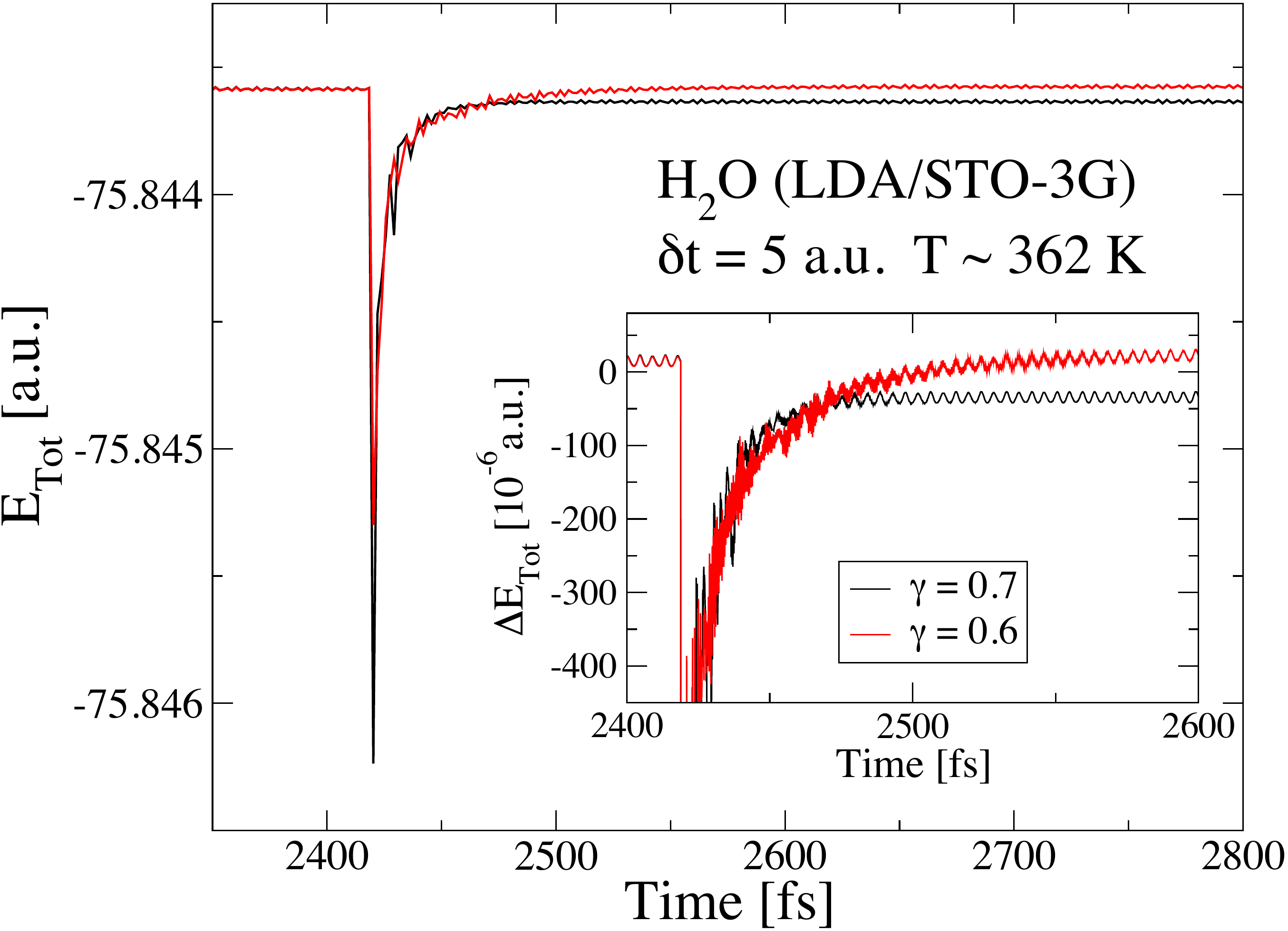}
\caption{\label{Figure_7}\small
Total energy of a fast-QMMD calculation of a single water molecule using the LDA density functional \cite{LDA}.
After $\sim$ 2400 fs the dynamics is perturbed by replacing the density matrix, $D$, with the initial density matrix, $D(t=0)$, calculated at the time, t = 0 s.
The perturbation calculation was performed for three different $\kappa$-parameters (Eq. (\ref{VRL_Damp})). %Here $E_{0}$ = -75.8436 a.u. a.u. 
All calculations were performed with the modified Verlet integration ($K$ = 7 in table \ref{Tab_Coef} ). 
Here rescaling of a $\kappa = \delta t^2 \omega^2$ with a factor $\gamma$ = 0.7
was employed for the black curve, and  $\gamma$ = 0.6 for the red curve.}
\end{figure}

The fast-QMMD scheme is robust to sudden perturbations. Figure \ref{Figure_7}
shows the total energy of a simulated water molecule. After about 2.4 ps of simulation
time an abrupt significant change in the density matrix $D(P)$ is introduced. Despite this sudden
large perturbation the energy relaxes and returns to a stable average value that is slightly shifted
but close to the initial total energy. However, a fairly short integration temp step is required to account
for the rapid change in density matrix.
We believe this response to a sudden perturbation
illustrates a key feature of the fast-QMMD scheme. Instead of optimizing the electronic
structure in each individual time step prior to the force calculation, the electronic
degrees of freedom is optimized dynamically as the trajectories propagate. The speed 
of the relaxation is slightly different for different choices of the scaling parameter $\gamma$.
This difference indicates an interesting opportunity. It should be possible to dynamically
update an optimal value of $\gamma$ in each time step. A time-dependent scaling 
factor, $\gamma(t)$, that is updated one-the-fly in a time-reversible way is 
an interesting opportunity for future developments. 

\section{Summary}

The goal of this paper was to explore the
extended Lagrangian formulation of Born-Oppenheimer molecular dynamics
in the limit of vanishing self-consistent field optimization.
In contrast to the most recent studies, Refs. \cite{PSouvatzis13}, we have generalized 
the first principles theory beyond the ground state Hartree-Fock method 
to include also free energy potential surfaces valid at finite electronic temperatures,
density functional theory with hybrid functionals, as well as 
the requirement of only one single Hamiltonian construction per time step.
This provides a very efficient and computational fast approach to first principles 
simulations that should be applicable to a broad class of materials.
Under normal conditions the proposed fast-QMMD scheme yields
trajectories that are practically indistinguishable from an "exact" Born-Oppenheimer
molecular dynamics simulation. However, even in the event of anomalous behavior that may
cause numerical instabilities, the optimization-free limit represents an ideal framework
for more elaborate force calculations that require an
improved accuracy in the electronic ground state optimization or a reduced 
length of the integration time step to recover stability or to improve accuracy.  

The optimization-free limit of extended Lagrangian Born-Oppenheimer molecular 
dynamics demonstrates some of the opportunities in the development of a
new generation first principles molecular dynamics 
that avoids current problems and shortcomings and allows 
a wider range of applications. Our work presented in this article 
is a step in this direction.

\section{Acknowledgements}

P. S. wants to thank L. C. for her eternal patience.
A.M.N.N acknowledge support by the United States Department of Energy (U.S. DOE) Office
of Basic Energy Sciences, discussions with M. Cawkwell, E. Chisolm, C.J. Tymczak, G. Zheng and stimulating contributions
by T. Peery at the T-Division Ten Bar Java group.  LANL is operated by Los Alamos National Security, LLC,
for the NNSA of the U.S. DOE under Contract No. DE-AC52- 06NA25396. Support by the G\"{o}ran Gustafsson
research foundation is also gratefully acknowledged.

\section{Appendix}

\subsection{Forces without prior self-consistent field optimization}

To derive the force expression in Eq.\ (\ref{FreeForce}), which together with the energy expression in Eq.\ (\ref{FreeE})
is one of our key results, we start by noting that
\begin{equation}\label{Taylor_1}\begin{array}{l}
E^{xc}[2D] = E^{xc}[2P] + 2Tr[(D-P)V^{xc}[2P]] \\
~\\
+ {\cal O}[(D-P)^2].
\end{array}
\end{equation}
Within the same order of accuracy, ${\cal O}((D-P)^2)$, 
we can replace the expression in Eq.\ (\ref{FreeE}) by
\begin{equation}\begin{array}{l}
{\cal F}[P] = 2Tr[hD] + Tr[(2D-P)G^\alpha (P)] \\
~\\
+ E^{xc}[2P] + 2Tr[(D-P)V_{xc}[2P]] \\
~\\
- T_e{\cal S}[D].
\end{array}
\end{equation}
Taking the partial derivative with respect to the nuclear coordinates (keeping $P$ constant) of the resulting free energy we get
\begin{equation}\label{ForceDeriv}\begin{array}{l}
 \left.\partial {\cal F}[P]/\partial R_I\right|_P  = {\cal F}_R = 2Tr[h_RD+hD_R] \\
~\\
~ + 2Tr[D_RG^\alpha (P)] + Tr[(2D-P)G^\alpha_R(P)] \\
~  \\
~ + E^{xc}_R[2P] + 2Tr[(D-P)V^{xc}_R(2P)] \\
~\\
~ + 2TR[D_RV^{xc}(2P)] - T_e\partial {\cal S}/\partial R_I \\
~~  \\
= 2Tr[h_RD] + Tr[(2D-P)G^\alpha_R(P)] \\
  \\
~ + E^{xc}_R[2P] + 2Tr[(D-P)V^{xc}_R(2P)] \\
~\\
~ + 2Tr[(h+G^\alpha (P) +V^{xc}(2P))D_R]  - T_e{\cal S}_R \\
~  \\
= 2Tr[h_RD] + Tr[(2D-P)G^\alpha_R(P)] \\
~  \\
~ + E^{xc}_R[2P] + 2Tr[(D-P)V^{xc}_R(2P)] \\
~\\
~  + 2Tr[H(P)D_R] - T_e{\cal S}_R \\
  \\
= 2Tr[h_RD] + Tr[(2D-P)G^\alpha_R(P)] \\
~\\
~ + E^{xc}_R[2D] + 2Tr[H(P)D_R] - T_e{\cal S}_R \\
~\\
~ + {\cal O}[(D-P)^2].
\end{array}
\end{equation}
In the last step  we used a second order approximation of $E^{xc}_R[2D]$ expanded 
around $2P$, i.e. as in Eq.\ (\ref{Taylor_1}), where
\begin{equation}\begin{array}{l}
E^{xc}_R[2D] = E^{xc}_R[2P] + 2Tr[(D-P)V^{xc}_R[2P]] \\
~~\\
 + {\cal O}[(D-P)^2].
\end{array}
\end{equation}
Here the subscript $R$ in $V^{xc}_R(2P)$, $E^{xc}_R[2P]$, $E^{xc}_R[2D]$ and $G^\alpha_R(P)$
denotes a derivative with respect only to the underlying atom centered basis.
We now need to simplify the
last two terms of the derivative above. We start with 
\begin{equation}\begin{array}{l}
2Tr[H(P)D_R] = 2Tr[H(\partial D/\partial R)] \\
~\\
= 2Tr[H\partial(ZD^\perp Z^T)/\partial R]\\
 ~ \\
 = 2Tr[H(Z_RD^\perp Z^T + Z D^\perp_R Z^T + ZD^\perp Z^T_R)] .
\end{array}
\end{equation}
By using the relation $Z_R = -(1/2)S^{-1}S_RZ$ proposed in Ref. \cite{ANiklasson07} we get
\begin{equation}\begin{array}{l}
 2Tr[H(P)D_R] = -Tr[HS^{-1}S_RD + HDS_RS^{-1}]\\
~\\
~ + 2Tr[H^\perp D^\perp_R]  \\
 ~ \\
 = -2Tr[S^{-1}HDS_R] + 2Tr[H^\perp D^\perp_R] , \\
\end{array}
\end{equation}
where we in the first step used the cyclic invariance under the trace, and in the next step
the commutation relation, $DH(P)S^{-1}-S^{-1}H(P)D = 0$. The last term above, $2Tr[H^\perp D^\perp_R]$, 
vanish only for idempotent solutions at zero electronic temperatures. At finite temperatures, however, the
$2Tr[H^\perp D^\perp_R]$ term is exactly cancelled by the entropy contribution, since
\begin{equation}\begin{array}{l}
{\displaystyle  T_e {\cal S}_R = -2T_ek_B\frac{\partial}{\partial R}Tr\left[D^\perp\ln(D^\perp ) \right.}\\
~\\
~ \left.+ (I-D^\perp)\ln(I-D^\perp )\right]  \\
 ~ \\
  = -2\beta^{-1}Tr[D^\perp _R\ln(D^\perp / (I-D^\perp )] \\
~\\
= -2 \beta^{-1}Tr[D^\perp _R\ln\left(e^{-\beta (H^\perp (P) - \mu I)}\right)]  \\
 ~ \\
  = 2Tr[D^\perp_R H^\perp ].
\end{array}
\end{equation}
In the last equation above we used the fact that $Tr[D^\perp_R] = 0$, since we assume canonical, charge conserving, 
partial derivatives. Going back to the force derivation in Eq.\ (\ref{ForceDeriv}) we now have that 
\begin{equation}\begin{array}{l}
{\cal F}_R[P] = 2Tr[h_R] + Tr[(2D-P)G^\alpha_R(P)] + E^{xc}_R[2D] \\
~\\
~ -2Tr[S^{-1}HDS_R] + {\cal O}[(D-P)^2]
\end{array}
\end{equation}
which completes our derivation of Eq.\ (\ref{FreeForce}).

\newpage
~

\end{document}